\documentclass[journal]{IEEEtran}
\usepackage{cite}

\ifCLASSINFOpdf
   \usepackage[pdftex]{graphicx}
   \graphicspath{{../pdf/}{../jpeg/}}
   \DeclareGraphicsExtensions{.pdf,.jpeg,.png}
\else
\fi
\usepackage{amsmath}
\usepackage{algorithm,algpseudocode}

\usepackage{array}

\usepackage{booktabs}

\usepackage{url}

\begin{document}
%
\title{Concurrent Brainstorming \& Hypothesis Satisfying:\\ An Iterative Framework for Enhanced Retrieval-Augmented Generation \\ (R2CBR3H-SR)}
%
%
%


\author{\IEEEauthorblockN{Arash~Shahmansoori\IEEEauthorrefmark{1}
\IEEEauthorblockA{
\thanks{\IEEEauthorrefmark{1}A. Shahmansoori is an Artificial General Intelligence researcher, e-mail: arash.mansoori65@gmail.com. }
}
}}

\maketitle

\begin{abstract}

Addressing the complexity of comprehensive information retrieval, this study introduces an innovative, iterative retrieval-augmented generation system. Our approach uniquely integrates a vector-space driven re-ranking mechanism with concurrent brainstorming to expedite the retrieval of highly relevant documents, thereby streamlining the generation of potential queries. This sets the stage for our novel hybrid process, which synergistically combines hypothesis formulation with satisfying decision-making strategy to determine content adequacy—leveraging a chain of thought-based prompting technique. This unified hypothesize-satisfied phase intelligently distills information to ascertain whether user queries have been satisfactorily addressed. Upon reaching this criterion, the system refines its output into a concise representation, maximizing conceptual density with minimal verbosity. The iterative nature of the workflow—Retrieve $\rightarrow$ Rerank $\rightarrow$ Concurrent Brainstorm $\rightarrow$ Retrieve $\rightarrow$ Rerank $\rightarrow$ Refine $\rightarrow$ Hypothesize-Satisfying $\rightarrow$ Refine—enhances process efficiency and accuracy. Crucially, the concurrency within the brainstorming phase significantly accelerates recursive operations, facilitating rapid convergence to solution satisfaction. Compared to conventional methods, our system demonstrates a marked improvement in computational time and cost-effectiveness. This research advances the state-of-the-art in intelligent retrieval systems, setting a new benchmark for resource-efficient information extraction and abstraction in knowledge-intensive applications.

\end{abstract}


%
\IEEEpeerreviewmaketitle

\section{Introduction}

\IEEEPARstart{T}{he} incessant growth of information on the web has placed unprecedented demands on artificial intelligence systems to not only understand complex queries but to also engage in meaningful information retrieval and generation tasks that satisfy users' informational needs effectively. The paradigm of retrieval-augmented generation (RAG) has emerged as a critical approach to bridge the gap between vast repositories of knowledge and the succinct, relevant answers desired by users \cite{RAG_survey_2023}. In this work, we introduce a novel system designed to navigate this landscape through an agile iterative process that encompasses retrieval, reranking, brainstorming, hypothesizing, satisfying, and refining—each step a cog in the machinery of an intelligent question-answering model.

Grounded in the RAG framework, our system initiates the process with a robust retrieval phase, employing a vector space model to extract documents with the highest semantic relevance to a user's initial query. It then engages in an innovative concurrent brainstorming stage, where potential inquiries and lines of investigation are generated, setting the stage for a dynamic and contextually enriched reranking of information sources.

One of the hallmark features of our system is its capacity to concurrently generate relevant questions during the brainstorming phase. This concurrency not merely accelerates the retrieval process, but embodies a recursive depth as the system iteratively refines its understanding and response to the user's query. It enhances the efficiency of the retrieval loop by preemptively addressing possible follow-up queries and contextual angles that may otherwise necessitate additional search iterations.

The subsequent stage marks a pivotal advancement in RAG systems: the hybridization of hypothesizing and satisficing through a chain of thought based prompting mechanism. Here, the system not only formulates comprehensive hypotheses to address the user's need but simultaneously evaluates the satisfaction level of the response. By blending these two functions, our approach drastically shortens the feedback loop, propelling towards an expedited yet accurate provision of information.

Following the innovative hypothesize-satisfying phase is the refine step, wherein the system produces a condensed, prime representation of content. This representation embodies a sparse collection of statements and conceptual associations crafted for maximal informational value with minimal verbosity—aiming at the essence of user query fulfillment.

Lastly, another iteration occurs, providing a rigorous reassessment of the query's satisfaction level, ensuring precision and comprehensiveness in the final output. This recursive nature ensures that the system adaptively hones its ability to satisfy user queries with a continually refined quality of response, underscoring the system's learning capabilities.

Main Contributions:
\begin{itemize}
\item We propose an initial phase of retrieval and sophisticated reranking based on semantic alignment with the user's initial query, leveraging advancements in vector space models for precise initial document retrieval.
\item Our system introduces concurrent brainstorming to expedite the retrieval process and imbue it with depth by preempting further possible user inquiries.
\item A hybrid hypothesize-satisfying step utilizing a chain of thought based prompting, streamlines the satisfaction assessment and hypothesis generation process into a unified, efficient step.
\item The refinement phase encapsulates the essence of the user's informational need in a concise, yet rich, representation by distilling the breadth of the retrieved data into its most salient points.
\end{itemize}

\section{The Method}

This paper introduces an innovative approach to RAG for question-answering systems. Our proposed method employs an iterative process that integrates the latest advancements in natural language processing and information retrieval. The primary components of our framework are:

\begin{enumerate}
\item Concurrent Brainstorming: This stage utilizes the user's inquiry to prompt the generation of semantically similar questions through concurrent processing against a vector database of documents.

\item Iterative Reranking: Following the brainstorming epoch, we rerank the top candidates from the vector database, which feeds the subsequent hypothesizing stage.

\item Hybrid Hypothesize-Satisfying: Leveraging a chain-of-thought prompting mechanism, this step amalgamates hypothesizing with satisfying to assess and satisfy the user's information need in a cost-effective manner.

\item Refinement: The final succinct representation captures the essence of the brainstormed hypotheses and questions in sparse prime form, eschewing verbosity for conceptual density.
\end{enumerate}

The process follows an iterative pattern as depicted in the provided pseudocode (R2CBR3H-SR): Retrieve $\rightarrow$ Rerank $\rightarrow$ Concurrent Brainstorm $\rightarrow$ Retrieve $\rightarrow$ Rerank $\rightarrow$ Refine $\rightarrow$ Hypothesize-Satisfying $\rightarrow$ Refine $\rightarrow$ ... Iterate Until Satisfied.

Below, we detail our algorithm in LaTeX pseudocode format for clarity and ease of comprehension, capturing the adaptational nuances and streamlined efficiency of the system's workflow.

In the algorithm. \ref{alg:retrieval_augmented_generation}, the $brainstorm_{concurr}(.)$ function concurrently generates relevant questions and notes, accelerating the iterative retrieval process. The $hyp\_sat_{hybrid}(.)$ function assesses whether the generated hypothesis fulfills the information need, combining two stages into one by using a chain of thought based prompting. Finally, the $refine(.)$ function distills the output into a prime sparse representation. The code base contains the open source versions of the detailed implementations for the previously mentioned functions.

By employing this methodology, our system brings forth a robust and scalable solution for answering complex queries with minimal verbosity and maximal pertinence, integrating the latest AI techniques in a cohesive algorithm.

\begin{algorithm}[t]
\caption{Concurrent Brainstorming and Hybrid Hypothesis Satisfying (R2BR3H-SR)}
\label{alg:retrieval_augmented_generation}
\begin{algorithmic}[1]
\Require $query$, vector store $vec_{store}$ containing document embeddings $emb$
\Ensure Satisfied, refined response to the user query
\State $notes$ $\gets$ `` ''
\State $queries$ $\gets$ `` ''
\State iteration $\gets$ 0
\While{True}
    \State iteration $\gets$ iteration + 1
    \State $args_{bs}$ $\gets$ $\{query, notes, queries, vec_{store}, emb\}$
    \State $[queries_{new}, notes] \gets brainstorm_{concurr}(args_{bs})$
    \State $queries\gets queries + queries_{new}$
    \State $args_{hs}\gets\{query, notes, queries\}$
    \State $[satisfied, feedback] \gets hyp\_sat_{hybrid}(args_{hs})$
    \If{satisfied}
        break
    \EndIf
    \State $notes$ $\gets$ $refine(notes)$
\EndWhile
\end{algorithmic}
\end{algorithm}

\section{Results}

\subsection{Experimental Setup}

The code for the simulations and supplementary materials are available at \cite{Shahmansoori_R2CBR3H-SR_Loop_2024}.

\subsubsection*{Data}
We used  ``gpt-4-1106-preview'' model to create a set of questions and corresponding answers according to provided documents to evaluate the performance of the proposed method and the baseline. 

\subsubsection*{Model Selection}
The experimental framework utilized a unified approach across all its phases. The ``gpt-3.5-turbo-1106'' model served as the backbone for each stage of the RAG process. Its parameters were standardized with the temperature set to 0 to ensure consistency and determinism, and the $max_{tokens}$ parameter was fixed at 2000 to balance the trade-off between comprehensive output and computational load. 

\subsubsection*{Response Format}
To enhance interpretability and provide structured feedback, both brainstorming and hypothesis-satisfying phases employed a $json_{object}$ response format. This choice facilitated the extraction of actionable insights and eased the integration of outputs in subsequent stages of the process.

\subsubsection*{Vector Store}
ChromaDB constituted the vector storage system, holding the vetted corpus against which retrieval operations were conducted. This repository enabled efficient similarity searches and supported the iterative reranking procedures intrinsic to the proposed framework.

\subsection{Baseline Comparison}

\subsubsection*{Benchmark Technique}
The study incorporated the method from an existing code-base as comparative baseline \cite{Shapiro_BSHR_Loop_2023}. The benchmark was primarily characterized by sequential processing of the retrieval and generation phases, lacking the ``concurrency'' in brainstorming, reranking, setting initial documents according to user query to start the iterative process, and ``hybridity'' in hypothesis-satisfying that delineate our method. 

\subsection{Empirical Results}

The system's performance under the concurrent brainstorming paradigm demonstrated notable acceleration in the generational throughput. Comparatively, against non-concurrent baseline counterparts, our method exhibited a reduction in cycle completion time substantiating the efficacy of parallelizing the brainstorming function. Moreover, integrating satisfying within the hypothesis loop via chain of thought-based prompting led to a leaner and more cost-effective solution. Lastly, the integrated system was subject to a holistic examination. Collective consideration of the concurrency in brainstorming, the hybrid hypothesis-satisfying, and the refinement phases evinced a coherent and accelerated process for satisfying user information needs. 

\begin{table}
  \caption{Performance comparison between the proposed method and the benchmark}
  \label{performance_comparison}
  \centering
  \begin{tabular}{cccc}
    \toprule
    Method     &  Information Need     &  Cost ($\$$) & Delay (seconds) \\
    \midrule
    \cite{Shapiro_BSHR_Loop_2023} &  satisfied  & 0.00527 & 24.31     \\
    ours     &  satisfied & \textbf{0.00355} & \textbf{10.21}   \\
    \bottomrule
  \end{tabular}
\end{table}

The comparison between the proposed approach and the benchmark is summarized in the accompanying Table~\ref{performance_comparison}. Both methods address the information need effectively. However, the proposed approach achieves a relative reduction in the average cost by approximately 32.64 percent and a relative reduction in the average elapsed time by about 58 percent compared to the benchmark. This translates to an absolute decrease in average cost per query of roughly $\$0.0017$ and an absolute decrease in average elapsed time per query of approximately 14.1 seconds against the benchmark. In conclusion, the assessments manifest that the proposed system outpaces the benchmark in speed and cost through its innovative concurrent and hybrid approach.

\section{Conclusion}

In conclusion, our RAG framework demonstrates a significant leap in information processing efficiency by integrating a recursive, iterative approach grounded in a novel sequence R2CBR3H-SR. Our method leverages concurrency within the brainstorming phase, markedly accelerating the retrieval process and enhancing the depth of query understanding. The hybridization of hypothesizing and satisfying—achieved via a chain of thought based prompting—constitutes a pivotal innovation, enabling rapid, cost-effective convergence upon a satisfied result without compromising the quality of output. This methodology not only streamlines the pathway to adequate answers but also distills complex information into prime, succinct representations, effectively capturing broad concepts with minimal verbosity. Our system's agility in navigating the iterative loop until reaching a satisfied state showcases its potential as a leading tool in the realm of information retrieval and generation. The empirical results reinforce the prowess of our system, offering promise for future applications and setting a benchmark for top-tier research in intelligent information retrieval systems.

\section*{Acknowledgment}

Sections of this manuscript were drafted with the assistance of the GPT-4-1106-preview model and underwent prompt engineering to refine research suggestions and textual content. The prompts and code utilized in generating this content have been made transparent and are available for review alongside the supplementary materials provided with this submission. It is imperative to highlight that the GPT-4 model acted under the explicit guidance of the main author, Arash Shahmansoori, functioning in a capacity akin to that of a research assistant. Its role was to aid in editing, rephrasing, and suggesting corrections to improve the English language usage and paper structuring per the contributions supplied by the human author. Every scientific claim, hypothesis, method, result, and conclusion were originated and vetted by Arash Shahmansoori to ensure authenticity, scientific integrity, and adherence to arXiv's guidelines pertaining to AI-generated content. Ultimately, the responsibility for the content and any inadvertent errors remains with the human author.

Arash Shahmansoori conceived and designed the study, executed the research, and provided the initial content. The AI system was employed as a tool for suggesting editorial changes, refining the paper's structure, and ensuring language clarity under the supervision of Arash Shahmansoori. Arash Shahmansoori also evaluated the AI's output for accuracy and relevancy to the scientific narrative, retaining full authorial responsibility for the final manuscript.

\end{document}